	%

\def\rpb{{R.~P.~Brent}} %
\def\awb{{A.~W.~Bojanczyk}}
\def\htk{{H.~T.~Kung}}
\def\ftl{{F.~T.~Luk}}

\def\IEEETC{{\it IEEE Transactions on Computers}}
\def\JACM{{\it J.~ACM}}

\def\SPIE{{Society of Photo-Optical Instrumentation Engineers}} %
\def\SISSC{{\it SIAM J.~Scientific and Statistical Computing}}

\magnification=\magstep1

\def\r{}			%

\hyphenpenalty=100
\vsize=20cm	%
\hsize=13cm
\baselineskip=12pt
\font\smallheadfont=cmr10 at 10truept
 at 10truept
\font\smallitfont=cmti10 at 10truept
\font\medbf=cmbx10 at 12.0pt

\settabs 6 \columns
\parindent 20pt
~
\bigskip
\centerline{\medbf Parallel Algorithms in Linear Algebra%
\footnote{}{\hskip-20pt\smallheadfont{Invited paper, to appear in
	{\smallitfont Proceedings Second NEC Research Symposium}
	(held at Tsukuba, Japan, August 1991).}}}
\bigskip
\centerline{Richard P.~Brent%
\footnote{}{\hskip-20pt\smallheadfont{Copyright \copyright\ 1991, R.~P.~Brent.
	    \hfill rpb128tr typeset using \TeX}}}
\centerline{Computer Sciences Laboratory}
\centerline{Australian National University}
\centerline{Canberra, Australia}
\centerline{\tt rpb@cslab.anu.edu.au}
\bigskip
\centerline{Report TR-CS-91-06}
\centerline{August 1991}
\footline={}				%

\bigskip
\leftline{\medbf Abstract}
\medskip

This paper provides an introduction to
algorithms for fundamental linear algebra problems
on various parallel computer architectures, with the emphasis
on distributed-memory MIMD machines.
To illustrate the basic concepts and key issues, we consider the
problem of parallel solution of a nonsingular linear system
by Gaussian elimination with partial pivoting.
This problem has come to be regarded as a benchmark for the performance
of parallel machines.  We consider its appropriateness as a benchmark,
its communication requirements, and schemes for data distribution
to facilitate communication and load balancing.
In addition, we describe some parallel algorithms for orthogonal (QR)
factorization and the singular value decomposition (SVD).

\bigskip
\leftline{\medbf 1.~Introduction~-- Gaussian elimination as a benchmark}
\medskip

Conventional benchmarks are often inappropriate for parallel machines.
A good benchmark needs to be a well-defined problem with a verifiable
solution, as well as being representative of the problems
which are of interest to users of the machine.
The problem should be scalable because the power of the
machine may be wasted on problems which are too small.
\medskip

For these reasons, the solution of a system of $n$ nonsingular linear
equations in $n$ unknowns, by the method of Gaussian elimination with
partial pivoting, has become a popular benchmark [\r15].
For conventional serial machines we can use $n = 100$, but for more powerful
machines $n$ can be increased to 1000 or more.
\medskip

In Section 2 we introduce some basic concepts such as
{\it speedup} and {\it efficiency} of parallel algorithms,
and {\it virtual processors}.
Various parallel computer architectures are outlined in Section 3.
\medskip

On parallel machines with distributed memory, questions of
data distribution and data movement are very important.
In deciding how to partition
data over the processors of a distributed-memory machine, attention must be
paid both to the data distribution patterns implicit in the algorithm
and the need to balance the load on the different processors.
In Section 4 we consider the data movement required for Gaussian
elimination, and how this maps to the movement of data between
processors.
This allows us to reach some conclusions about the appropriateness of
different machine architectures for linear algebra computations.
\medskip

To illustrate the key issues, Section 5 considers in more detail
the problem of parallel solution of a nonsingular linear system by
Gaussian elimination with partial pivoting on a distributed-memory
MIMD machine.
\medskip

\footline={\hss\tenrm\folio\hss}	%
Because of the difficulties and communication overheads associated with
pivoting, it is tempting to try to avoid pivoting,
but omission of pivoting in Gaussian elimination leads to
numerical instability.  One solution is to implement a parallel
version of the orthogonal (QR) decomposition instead of the
triangular (LU) decomposition obtained by Gaussian elimination.
This permits a stable solution without pivoting, but at the
expense of an increase in the number of floating-point operations.
The QR decomposition has many other useful applications,
e.g. to the solution of linear least squares problems or as
a preliminary step in the singular-value decomposition.
Some ways of implementing the QR decomposition in parallel
are mentioned in Section 5.3.
\medskip

Many problems in numerical linear algebra are easy to solve if we
can find the singular value decomposition (SVD) of a rectangular
matrix, or the eigenvalues and eigenvectors of a symmetric
(or Hermitian) matrix.  In Section 6 we describe some good parallel
algorithms for these problems. Often the parallel
algorithms are not just 
a straightforward modification of the best serial algorithms.
\medskip

There has been an
explosive growth of interest in parallel algorithms (including those
for linear algebra problems) in recent years, so we can not attempt to
be comprehensive.  For more detailed
discussions and additional references, the reader is referred to
surveys such as those by Gallivan {\it et al} [\r23],
Dongarra {\it et al} [\r16], and Heller~[\r36].

\bigskip
\leftline{\medbf 2.~Basic concepts}
\medskip

We assume that a parallel machine with $P$ processors is available.
Thus $P$ measures the degree of parallelism; $P = 1$ is just the familiar
serial case.  When considering the solution of a particular problem,
we let $T_P$ denote the time required to solve the problem using (at most)
$P$ processors.  The {\it speedup} $S_P$ is defined by
$$S_P = T_1/T_P,$$
and the {\it efficiency} $E_P = S_P/P$.
\medskip

When converting a serial algorithm into a parallel algorithm, our aim is
usually to attain constant efficiency,
i.e. $$E_P \ge c$$
for some positive constant $c$ independent of $P$.  This may be written as
$$E_P = \Omega(1).$$
Equivalently, we want to attain {linear} speedup, i.e. $$S_P \ge cP,$$
which may be written as $$S_P = \Omega(P).$$

\medskip
\leftline{\bf 2.1 Amdahl's Law}
\medskip

Suppose a positive fraction $f$ of a computation is ``essentially serial'',
i.e. not amenable to any speedup on a parallel machine.
Then we would expect
$$T_P = fT_1 + (1 - f)T_1/P$$
so the overall speedup
$$S_P = {1 \over f + (1 - f)/P} \le {1 \over f}, \eqno(2.1)$$
i.e. the speedup is {\it bounded}, not linear.  The inequality (2.1) is called
{\it Amdahl's Law} [\r1]
and has been used as an argument against parallel
computation.  However, what it shows is that the speedup is bounded as
we increase the number of processors for a {\it fixed} problem.  In practice,
it is more likely that we want to solve larger problems as the
number of processors increases, because the desire to solve larger problems is
a primary motivation for building larger parallel machines.
\medskip

Let $N$ be a measure of the problem size.  For many problems it is reasonable
to assume that
$$f \le K/N \eqno(2.2)$$
for some constant K.  For example, in problems involving $N$ by $N$ matrices,
we may have $\Omega(N^3)$ arithmetic operations and $O(N^2)$ serial
input/output.
\medskip

Suppose also that $N$ increases at least linearly with $P$,
with the same constant as in (2.2),
i.e. $$N \ge KP. \eqno(2.3)$$
(2.2) and (2.3) imply that $fP \le 1,$ so from (2.1) we have
$$S_P = {P \over fP + (1 - f)} \ge {P \over 2 - f} \ge {P \over 2}.$$
Thus we get linear speedup, with efficiency $E_P \ge 1/2$.
\medskip

For further discussion of Amdahl's law and the scaling of problem size,
see [\r33].

\medskip
\leftline{\bf 2.2 Virtual processors}
\medskip

In practice any parallel machine has a fixed maximum number (say $P$)
of processors
imposed by hardware constraints.  When analysing parallel algorithms
it is often convenient to ignore this fact and assume that we have as
many (say $p$) processors as desired.  We may think of these $p$ processors
as {\it virtual processors} or {\it processes}.  Each real processor
can simulate $\lceil p/P \rceil$ virtual processors (provided the real
processor has enough memory).  Thus, ignoring overheads associated with
the simulation, we have
$$S_P \ge S_p/{\lceil p/P \rceil}. \eqno(2.4)$$
If our analysis for $p$ processors gives us a lower bound on $S_p$, then
(2.4) can be used to obtain a lower bound on $S_P$.
In practice this may be an oversimplification, because the
hardware/software system may or may not provide good support for virtual
processors.

\bigskip
\leftline{\medbf 3. Parallel architectures}
\medskip

Many varieties of parallel computer architecture have been proposed in recent
years.  They include~--

\smallskip
\item{$\cdot$} Pipelined vector processors such as the Cray~1,
	Fujitsu VP~100, or NEC~SX/2,
	in which there is
	a single instruction stream and the parallelism is more or less
	hidden from the programmer.
	(More recent descendants such as the Cray~2S, Cray~Y-MP/8,
	and Fujitsu VP~2000 series incorporate parallel vector processors.)
\smallskip

\item{$\cdot$} Single-instruction multiple-data (SIMD [\r19]) machines such as
	the Illiac IV, ICL DAP, or MasPar MP-1,		%
        in which a number of simple processing elements
	(PEs or cells) execute the same instruction on local data
	and communicate with their nearest neighbours on a square grid
	or torus.  There is usually a general-purpose controller which
	can broadcast instructions and data to the cells.
\smallskip

\item{$\cdot$} Multiple-instruction multiple-data (MIMD) machines such as
	those constructed from transputers,
	the Carnegie-Mellon CM*,
	the Fujitsu AP~1000,
	and hypercube machines such as the
	Caltech ``Cosmic Cube'', Intel iPSC, and nCUBE2.
	In the hypercube machines $2^k$ processors are connected like
	the vertices of a $k$-dimensional cube, i.e. the processors
	are identified by $k$-bit binary numbers, and are connected to
	the processors whose numbers differ by exactly one bit from
	their own.
\smallskip

\item{$\cdot$} Massively parallel SIMD machines such as the CM-200
	Connection Machine (which may also be regarded as a hypercube
	machine).
\smallskip

\item{$\cdot$} Shared-memory multiprocessors such as the Alliant FX/80,
	Cedar, Encore Multimax, and Sequent Symmetry.
\smallskip

\item{$\cdot$} Systolic arrays [\r41], which are 1 or 2-dimensional
	arrays of simple processors (cells) connected to their
	nearest neighbours.  The cells on the edge of the array are
	usually connected to a general-purpose machine which acts as a
	controller.  Examples are the Warp and iWarp machines [\r2, \r6]
	and several
	machines described in [\r53].   Variations on the idea of
	systolic arrays are wavefront arrays [\r45] and
	instruction systolic arrays [\r58].
\medskip

The categories listed above are not mutually exclusive. For example, there is
an overlap between vector processors and shared-memory multiprocessors.
Also, the distinction between SIMD and MIMD machines is orthogonal to the
distinction between hypercube, grid and torus topologies.
\medskip

In view of the diversity of parallel computer architectures, it is
difficult to describe practical parallel algorithms in a machine-independent
manner.  In some cases an algorithm intended for one class of parallel
machine can easily be converted for another (more general) class.
For example, an algorithm designed for a systolic array can easily be mapped
onto a hypercube, but not conversely
(in general).  An algorithm designed for a distributed-memory machine can
easily be implemented on a shared memory machine, but the
converse may not be true.  As a rough approximation, it is easier to
implement algorithms on the more general machines, but the more
specialised machines may be more efficient (or cost-effective) for
certain classes of problems.
For example, systolic arrays are sufficient
and cost-effective for many problems arising in digital signal
processing [\r7, \r31, \r41, \r44, \r45, \r53].
\medskip

In the following sections we describe algorithms for
distributed-memory message-passing machines or systolic arrays~-- the reader
should be able to translate to other appropriate architectures.

\bigskip
\leftline{\medbf 4. Data movement and data distribution}
\medskip

Before considering how to map data (e.g.~vectors and matrices) onto
distributed-memory machines, it is worth considering what forms of 
data movement are common in linear algebra algorithms.
For the sake of example we focus our attention on Gaussian elimination
with partial pivoting, but most other linear algebra algorithms have
similar data movement requirements.  To perform Gaussian elimination
we need~--
\smallskip
\item{$\cdot$} Row/column broadcast.  For example, the pivot row needs to
	be sent to processors responsible for other rows, so that they
	can be modified by the addition of a multiple of the pivot row.
	The column which defines the multipliers also needs to be
	broadcast.
\smallskip
\item{$\cdot$} Row/column send/receive.  For example, if pivoting is
	implemented by explicitly interchanging rows, then at each
	pivoting step two rows have to be interchanged.  (This could
	be done by broadcasting both rows, but it might
	be less efficient than explicitly sending the rows to the
	appropriate destinations.)
\smallskip
\item{$\cdot$} Row/column scan. Here we want to apply an associative
	operator $\theta$ to data in one (or more) rows or columns.
	For example, when selecting the pivot row it is necessary to
	find the index of the element of maximum absolute value in
	(part of) a column.  This may be computed via an
	associative operator $\theta$ defined on pairs:
	$$(a,i) \;\theta\; (b,j) = 
	  \cases{(a,i),&if $\vert a \vert \ge \vert b \vert$;\cr
	         (b,j),&otherwise.\cr} \eqno(4.1)$$
	Other useful associative operators are addition (of scalars or
	vectors) and concatenation (of vectors).

\medskip
\leftline{\bf 4.1 Data distribution}
\medskip

On a distributed-memory machine
where each processor has a local memory which is
accessible to other processors only by explicit
message passing, it is customary to partition
data such as matrices and vectors across the local memories of
several processors.  This is essential for problems which are too
large to fit in the memory of one processor, and
in any case it is usually desirable
for load-balancing reasons.  (The exception is for
very small problems which may as well be solved in a single processor.)
Since vectors are a special case of matrices, we consider the
partitioning of an $m$ by $n$ matrix $A$.
\medskip

It is desirable for data to be distributed in a ``natural'' manner, 
so that the operations of row/column broadcast/send/receive/scan
described above can be implemented efficiently.  This is possible if
a square grid is a subgraph of the connection graph of the parallel
machine.  For example, it is true for machines whose connection topology
is an $s$ by $s$ torus or a hypercube of even dimension.  
(On machines
for which a rectangular grid of moderate aspect ratio can be embedded,
say an $s$ by $ks$ grid for some small positive integer $k$, we can
use a square {\it virtual} $ks$ by $ks$ grid by having each processor
simulate $k$ virtual processors.)
\medskip

The simplest mapping of data to processors is the {\it column-wrapped}
(or {\it row-wrapped}) representation.  Here column (or row) $i$ of
a matrix is stored in the memory associated with processor
$i \bmod P$, assuming that the $P$ processors are numbered
$0, 1, \ldots, P-1$.  (A Fortran programmer might prefer
$i-1 \bmod P$, but we find that C array conventions are more convenient.)
\medskip

Although simple, and widely used in parallel implementations of
Gaussian elimination (e.g.~[\r24, \r25, \r46, \r52]),
the column-wrapped
(or row-wrapped) representation has some disadvantages~--
\smallskip
\item{$\cdot$} Lack of symmetry~-- rows are treated differently
from columns. It is instructive to consider the data communication
involved in transposing a matrix.
\smallskip
\item{$\cdot$} Poor load-balancing for moderate-sized problems~--
if $n < P$ some processors store no columns, so presumably perform no
useful work.  On the other hand, if $n$ increases from $P$ to $P+1$ then
the load on processor 0 doubles.  Thus, the performance curve as a function
of $n$ (the number of columns in the matrix) will be ``jagged''~-- there
will be jumps at each multiple of $P$.
\smallskip
\item{$\cdot$} Poor communication bandwidth for column broadcast~--
since each column is stored in the memory associated with only one
processor, the speed of column broadcast is constrained by the
communication bandwidth of a single processor (compare the blocked/scattered
storage representations below).
\medskip

Another conceptually simple mapping is the {\it blocked} representation.
Assume that the processors form an $s$ by $s$ grid ($P = s^2$).
The matrix $A$ is padded with zero rows and columns if necessary, so that
$m$ and $n$ can be assumed to be multiples of $s$.
$A$ is partitioned into an $s$ by $s$ matrix of blocks.
Each block, of dimension $m/s$ by $n/s$, is assigned to one processor
in the natural way. This avoids the lack of symmetry inherent in the
row/column-wrapped representation. It also
improves the communication bandwidth for column broadcast, because each
column is shared by $s = P^{1/2}$ processors.
However, it suffers from a load-balancing problem~--
\smallskip
\item{$\cdot$} Poor load balancing for triangular and band matrices~--
if $A$ is upper triangular then about half the processors (those
storing the strict lower triangle of blocks) are only storing zeros
and can probably not do any useful computation.  Similarly (but even worse)
if $A$ is a band matrix with bandwidth small relative to $m$ and~$n$.
\medskip

Harder to visualize, but often better than the row/column-wrapped or
blocked representations, is the {\it scattered} representation [\r21]
(also called {\it dot mode} in image-processing applications [\r39]).
Assume as above that the processors form an $s$ by $s$ grid,
and let the processors be numbered $(0, 0), (0, 1), \ldots, (s-1, s-1)$.
Then the matrix element $a_{i,j}$ is stored in processor
$(i \bmod s, j \bmod s)$.
Now the matrices stored locally on each processor have the same shape
(e.g. triangular, band, $\ldots$) as the global matrix $A$, so the
computational load on each processor is approximately equal.
\medskip

It is sometimes useful to regard a blocked representation of a matrix
as a scattered representation of the same matrix with its rows and 
columns permuted. Formally, if $s|k$,
define a $k$ by $k$ permutation matrix 
$\pi_k$ whose $(i,j)$-th element is 1 iff 
$$j = \cases{i.s \bmod k-1,&if $i < k-1$;\cr
	     k-1,&if $i = k-1$\cr} \eqno(4.2)$$
(assuming C conventions~-- indices run from 0 to $k-1$).
If $P = s^2$, $s|m$, $s|n$, then the scattered representation of
the $m$ by $n$ matrix $A$ is the same as the block representation
of $\pi_mA\pi_n^{-1}$.  Similarly, if $B$ is $n$ by $p$, $s|p$,
then the scattered representation of $B$ is the same as the block
representation of $\pi_nB\pi_p^{-1}$, and the scattered representation
of $AB$ is the same as the block representation of
$\pi_mAB\pi_p^{-1} = (\pi_mA\pi_n^{-1})(\pi_nB\pi_p^{-1})$. 
This shows formally that a matrix multiplication algorithm which works
for matrices stored using the blocked representation should also
work for matrices stored using the scattered representation, and
vice versa.
\medskip

The blocked and scattered representations do not actually require a
square processor array~-- rectangular would suffice.
The reason why we ask for the processor
array to be square is that this makes matrix multiplication and
matrix transposition much simpler than in the general case.

\medskip
\leftline{\bf 4.2 Implications for architectures}
\medskip

First consider distributed memory machines.
In order to have a natural mapping of global matrices to a scattered
storage representation, the inter-processor connection graph should
have a square grid as a subgraph.  A square torus is safisfactory,
though slightly more general than necessary.  A hypercube of even
dimension is also satisfactory, but considerably more general than
necessary.
\medskip

A hypercube does have some advantages over a simple square $s$ by $s$
grid.  Consider communication along one row (or column) of the grid.
With a simple grid the maximum distance between processors in one row is
$s-1$, so time required to send or broadcast a short message is of order $s$
in the worst (and average) case.  With a hypercube this is reduced to
order $\log s$.
\medskip

In practice, for small and moderate values of $s$, the reduction from
order $s$ to order $\log s$ may not be so important as the overhead
incurred at each intermediate node in the path from source to destination.
For example, early hypercube machines used {\it store and forward},
which imposed a considerable processing load on the intermediate
nodes.  Consider sending a message of length $n$.  At each intermediate
node the delay is of order $n$, so the overall delay is of order
$n\log s$.
\medskip

Some more recent hypercube and torus machines (e.g.~the Fujitsu AP~1000)
have used {\it wormhole routing} [\r14] rather than store and forward.
With wormhole routing the overall delay is
of order $n + \log s$ (for a hypercube) or $n + s$ (for a grid).
Also, the wormhole routing protocol can be implemented in hardware
so as not to impose a load on the processors at intermediate nodes.
\medskip

For shared memory machines the problems are different.
Clearly the memory must have sufficiently high overall bandwidth,
but this is not sufficient.  The problem of {\it hotspots},
i.e.~memory locations accessed intensively by several processors,
needs to be overcome.  For example, in Gaussian elimination the
memory locations where the pivot row is stored will be hotspots.
On a distributed memory machine the programmer (or compiler) has
to solve this problem by explicitly broadcasting the pivot row,
but on a shared memory machine the hardware needs to be able to
provide for simultaneous read-only access to the pivot row by
several processors.

\bigskip
\leftline{\medbf 5. The solution of linear systems}
\medskip

Suppose we want  to solve a nonsingular $n$ by $n$ linear system
$$Ax = b \eqno(5.1)$$
on a parallel machine for which a 2-dimensional mesh
is a natural interconnection pattern.  It is easy to implement Gaussian
elimination {\it without} pivoting, because multipliers can be propagated
along rows of the augmented matrix $[A \vert b]$, and it is not necessary for
one row operation to be completed before the next row operation starts.
Unfortunately, as is well-known [\r32, \r60, \r63],
Gaussian elimination without pivoting
is numerically unstable unless $A$ has some special property such as
diagonal dominance or positive definiteness. Thus we consider the 
implementation of Gaussian elimination with partial pivoting
on a parallel machine.

\medskip
\leftline{\bf 5.1 Gaussian elimination with partial pivoting}
\medskip

For the sake of definiteness, we assume that Gaussian elimination is to
be performed on a distributed-memory machine with an $s$ by $s$ grid,
and that the augmented matrix $[A \vert b]$ is stored in the scattered
representation.  The reader should be careful to distinguish between
a row (or column) of {\it processors} and a row (or column) of the
{\it matrix}. Generally, each row of processors stores several rows of
the matrix.
\medskip

It is known [\r32, \r60, \r63]
that Gaussian elimination is equivalent to triangular
factorization.  More precisely, Gaussian elimination with partial
pivoting produces an
upper triangular matrix $U$ and a lower triangular matrix $L$ (with unit
diagonal) such that 
$$PA = LU \eqno(5.2)$$
where $P$ is a permutation matrix (not the number of processors here !).  
In the usual implementation $A$
is overwritten by $L$ and $U$ (the diagonal of $L$ need not be stored).
If the same procedure is applied to the augmented matrix $\bar A = [A \vert b]$,
we obtain
$$P\bar A = L\bar U \eqno(5.3)$$
where $\bar U = [U \vert \bar b]$
and (5.1) has been transformed into the upper triangular system
$$Ux = \bar b \eqno(5.4)$$
In the following we shall only consider the transformation of $A$ to
$U$, as the transformation of $b$ to $\bar b$ is similar.
\medskip

If $A$ has $n$ rows, the following steps have to be repeated $n-1$ times,
where the $k$-th iteration completes computation of
the $k$-th column of $U$~--
\smallskip
\item{1.} Find the index of the next pivot row by finding an element of
maximal absolute value in the current ($k$-th) column,
considering only elements on and below the diagonal. 
With the scattered representation this
involves $s$ processors, which each have to find a local maximum and
then apply the associative operator (4.1).
\smallskip
\item{2.} Broadcast the pivot row vertically.
\smallskip
\item{3.} Exchange the pivot row with the 
current $k$-th row,
and keep a record of the row permutation.  Generally the exchange requires
communication between two rows of $s$ processors.  Since the pivot row
has been broadcast at step 3, only the current $k$-th row
needs to be sent at this step.  (Alternatively, the exchanges could
be kept implicit, but this would lead to load-balancing problems and
difficulties in implementing block updates, so explicit exchanges are
usually preferable).
\smallskip
\item{4.} Compute the ``multipliers'' (elements of $L$) from the
$k$-th column and broadcast horizontally.
\smallskip
\item{5.} Perform Gaussian elimination (using the portion of the pivot
row and the other rows held in each processor).  If done in the
obvious way, this involves saxpy operations
(a {\it saxpy} is the addition of a scalar multiple of one vector to
another vector),
but the computation can also be formulated as a rank-1 update.
\medskip

We can make an estimate of the parallel time $T_P$ required to perform
the transformation of $A$ to upper triangular form.
There are two main contributions~--
\smallskip
\item{A.} {\it Floating-point arithmetic.} The overall computation involves
$2n^3/3 + O(n^2)$ float\-ing-point operations (counting additions and
multiplications separately). Because of the scattered representation
each of the $P = s^2$ processors performs approximately the same
amount of arithmetic.  Thus floating-point arithmetic contributes a term
$O(n^3/{s^2})$ to the computation time.
\smallskip
\item{B.} {\it Communication.} At each iteration of steps 1-5 above,
a given processor sends or receives $O(n/s)$ words.  We shall assume that
the time required to send or receive a message of $w$ words is
$c_0 + c_1w$, where $c_0$ is a ``startup'' time and $1/c_1$ is the
transfer rate.  (For real machines the time may depend on other
factors, such as the distance between the sender and the receiver
and the overall load on the communication network.)
With our assumption, the overall communication time is $O(n^2/s) + O(n)$,
where the $O(n)$ term is due to startup costs.
\medskip

If arithmetic and communication can not be overlapped, the overall
time $T_P$ is simply the sum of A and B above, i.e.
$$T_P \simeq \alpha n^3/s^2 + \beta n^2/s + \gamma n, \eqno(5.5)$$
where $\alpha$ depends on the floating-point 
and memory speed of each processor,
$\beta$ depends mainly on the communication transfer rate between processors,
and $\gamma$ depends mainly on the communication startup time.
We would expect the time on a single processor to be
$$T_1 \simeq \alpha n^3, \eqno(5.6)$$
although this may be inaccurate for various reasons~-- e.g.\ the problem
may fit in memory caches on a parallel machine, but not on a single
processor.
\medskip

From (5.5) and (5.6), the efficiency $E_P$ is
$$E_P \simeq {1 \over 
	{1 + (1 + \bar\gamma/\bar n)\bar\beta/\bar n}}, \eqno(5.7)$$
where $\bar\beta = \beta/\alpha$ is proportional to the 
ratio of communication to computation speed,
$\bar\gamma = \gamma/\beta$ measures the importance of the
communication startup time,
and $\bar n = n/s$ is the number of rows or columns of $A$ stored
in a single processor. 
From (5.7), the efficiency is close to 1 only if $\bar n \gg \bar \beta$.
\medskip

We have ignored the ``back-substitution'' phase, i.e. the solution of
the upper triangular system (5.4), because this can be performed in time
much less than (5.5) (see [\r21, \r46]).

\medskip
\leftline{\bf 5.2 Blocking}
\medskip

On many machines it is impossible to achieve peak performance if
the Gaussian elimination is performed via saxpys or rank-1 updates.
This is because performance is limited by memory accesses rather than
by floating-point arithmetic, and saxpys or rank-1 updates have a
high ratio of memory references to floating-point operations.
Closer to peak performance can be obtained for matrix-vector or
(better) matrix-matrix multiplication.
\medskip

It is possible to reformulate Gaussian elimination so that most of
the floating-point arithmetic is performed in matrix-matrix
multiplications, without compromising the error analysis.
Partial pivoting introduces some difficulties, but they are
surmountable.  The idea is to introduce a ``blocksize'' or
``bandwidth'' parameter $\omega$.  Gaussian elimination is
performed via saxpys or rank-1 updates
in vertical strips of width $\omega$.  Once $\omega$ pivots have
been chosen, a horizontal strip of height $\omega$ can be updated.
At this point, a matrix-matrix multiplication can be used to update the lower
right corner of $A$.  The optimal choice of $\omega$ depends on
details of the machine architecture, but
$$\omega \sim n^{1/2}	\eqno(5.8)$$
is a reasonable choice.
\medskip

The effect of blocking is to reduce the constant $\alpha$ in (5.5)
at the expense of increasing the lower-order terms. Thus, a blocked
implementation should be faster for sufficiently large $n$,
but may be slower than an unblocked implementation for small~$n$.

\medskip
\leftline{\bf 5.3 Orthogonal factorization}
\medskip

On machines whose architecture makes pivoting difficult, we can avoid it
at the expense of increasing the amount of arithmetic.  For example,
on systolic arrays it is possible to compute the orthogonal
(QR) factorization of $A$ efficiently and in a numerically stable
manner using Givens transformations [\r5, \r27, \r47]. The cost is an
increase by a factor of four in the number of arithmetic operations
(though this factor may be reduced if ``fast'' Givens transformations
[\r26] are used).
In any case, the QR factorization is of independent interest because
it can be used to solve the linear least squares problem
$$\min \Vert Ax - b \Vert_2	\eqno(5.9)$$
where $A$ is an $m$ by $n$ matrix of rank $n$.
\medskip

On a single processor Householder transformations are cheaper than
Givens transformations. The steps involved in implementing a Householder
QR factorization on a parallel machine are similar to those involved in
implementing Gaussian elimination.  Although pivoting is not usually
required, the vectors $u$ which define the Householder transformations
$I - 2uu^T$ need to be broadcast in the same way as the pivot row and
multiplier column in Gaussian elimination.
\medskip

As for Gaussian elimination, optimal performance for large matrices may
require blocking.  Several Householder transformations can be combined
[\r3, \r57] and then applied together so that most of the arithmetic
is done in matrix-matrix multiplication.

\bigskip
\leftline{\medbf 6. The SVD and symmetric eigenvalue problems}
\medskip

A singular value decomposition (SVD) of a real $m$ by $n$ matrix $A$ is its
factorization into the product of three matrices:
$$A = U \Sigma V^T, \eqno(6.1)$$
where U is an $m$ by $n$ matrix with orthonormal columns, $\Sigma$ is an
$n$ by $n$ nonnegative diagonal matrix, and $V$ is an $n$ by $n$ orthogonal
matrix (we assume here that $m \ge n$).  The diagonal elements $\sigma_i$
of $\Sigma$ are the {\it singular values} of $A$.  The singular value
decomposition has many applications [\r29, \r44].
\medskip

The SVD is usually computed by a two-sided orthogonalization process,
e.g. by two-sided reduction to bidiagonal form 
(possibly preceded by a one-sided reduction [\r11]), followed by the
QR algorithm [\r28, \r30, \r63].
It is difficult to implement this Golub-Kahan-Reinsch
algorithm efficiently on a parallel
machine. It is much simpler (though perhaps less efficient) to use a one-sided
orthogonalization method due to Hestenes [\r37].
The idea is to generate an orthogonal matrix
$V$ such that $AV$ has orthogonal columns.  Normalizing the Euclidean length
of each nonnull column to unity, we get
$$AV = \tilde U \Sigma \eqno(6.2)$$
As a null column of $\tilde U$ is always associated with a zero diagonal
element of $\Sigma$, there is no essential difference between (6.1) and (6.2).
\medskip

The cost of simplicity is an increase in the operation count, compared to
the Golub-Kahan Reinsch algorithm.

\medskip
\leftline{\bf 6.1 Implementation of the Hestenes method}
\medskip

Let $A_1 = A$ and $V_1 = I$.  The Hestenes method uses a sequence of plane
rotations $Q_k$ chosen to orthogonalize two columns
in $A_{k+1} = A_kQ_k$.  If the matrix $V$ is required, the plane rotations
are accumulated using $V_{k+1} = V_kQ_k$.  Under certain conditions
$\lim Q_k = I$, $\lim V_k = V$ and $\lim A_k = AV$.
The matrix
$A_{k+1}$ differs from $A_k$ only in two columns, say columns $i$ and $j$.
In fact
$$\left(a_i^{(k+1)}, a_j^{(k+1)}\right) =
  \left(a_i^k, a_j^k\right) \left(\matrix{\cos \theta &\sin \theta\cr
             -\sin \theta &\cos \theta\cr}\right)$$
where the rotation angle $\theta$ is chosen so that the two new columns
$a_i^{(k+1)}$ and $a_j^{(k+1)}$ are orthogonal.  This can always be done
with an angle $\theta$ satisfying
$$\vert \theta \vert \le \pi/4, \eqno(6.3)$$
see for example [\r9].
\medskip

It is desirable for a ``sweep'' of $n(n-1)/2$ rotations to include all
pairs $(i,j)$ with $i < j$.  On a serial machine a simple strategy is to
choose the ``cyclic by rows'' ordering
$$ (1,2), (1,3), \cdots, (1,n), (2,3), \cdots, (n-1,n).$$
Forsythe and Henrici [\r20] have shown that the cyclic by rows ordering
and condition (6.3) ensure convergence of the Jacobi method applied to
$A^TA$, and convergence of the cyclic by rows Hestenes method follows.
In practice only a small number of sweeps are required.
The speed of convergence is discussed in [\r9].

\medskip
\leftline{\bf 6.2 The Go tournament analogy}
\medskip
 
On a parallel machine we would like to orthogonalize several pairs of
columns simultaneously.  This should be possible so long as no column
occurs in more than one pair.  The problem is similar to that of
organizing a round-robin tournament between $n$ players.
A game between players $i$ and $j$ corresponds to orthogonalizing
columns $i$ and $j$, a round of several games played at the same time
corresponds to orthogonalizing several pairs of (disjoint) columns,
and a tournament where each player plays each other player once
corresponds to a sweep in which each pair of columns is orthogonalized.
Thus, schemes which are well-known to Go players
(or players of other two-person games such
as Chess, Sumo, $\ldots$)
can be used to give orderings amenable to parallel computation.
It is usually desirable to minimize the number of parallel steps in a sweep,
which corresponds to the number of rounds in the tournament.
\medskip

On a parallel machine with restricted communication paths
there are constraints on the orderings which we can
implement efficiently.  A useful analogy is a tournament of lazy
Go players.  After each round the players want to walk only a short
distance to the board where they are to play the next round.
\medskip

Using this analogy, suppose that each Go board corresponds to a
virtual processor and each player corresponds to a column of the matrix
(initially $A$ but modified as the computation proceeds).  A game
between two players corresponds to orthogonalization of the corresponding
columns.  Thus we suppose that each virtual processor has sufficient
memory to store and update two columns of the matrix.  If the
Go boards (processors)
are arranged in a linear array with nearest-neighbour communication paths,
then the players should have to walk (at most) to an adjacent board
between the end of one round and the beginning of the next round,
i.e. columns of the matrix should have to be exchanged
only between adjacent processors.
Several orderings satisfying these conditions have
been proposed [\r8, \r9, \r50, \r56].
\medskip

Since $A$ has $n$ columns and at most $\lfloor n/2 \rfloor$ pairs can be
orthogonalized in parallel, a sweep requires as least $n-1$ parallel
steps ($n$ even) or $n$ parallel steps ($n$ odd).  The 
ordering of [\r9] attains this minimum, and convergence
can be guaranteed if $n$ is odd [\r50, \r59].  It is an open
question whether convergence can be guaranteed,
for any ordering which requires only the minimum
number of parallel steps, when $n$ is even~-- the problem in proving
convergence is illustrated in [\r34].
However, in practice lack of
convergence is not a problem, and it is easy to ensure convergence by
the use of a ``threshhold'' strategy [\r63],
or by taking one additional parallel step per sweep when $n$ is even [\r51].
\medskip

As described above, each virtual processor deals with two columns, so the
column-wrapped representation is convenient.  However, the block or
scattered representations can also be used. The block representation
involves less communication between real processors
than does the scattered representation if the
standard orderings are used.
However, the two representations are equivalent if different
orderings are used. 
(As in the discussion of matrix multiplication at the end of Section 4.1,
the algorithm can assume that the block representation is used, since the 
SVD of $\pi_mA\pi_n^{-1}$ is just a reordering of the SVD of $A$.)
The scattered representation does not have a load-balancing advantage
here, since the matrix does not change shape.

\medskip
\leftline{\bf 6.3 The symmetric eigenvalue problem}
\medskip

There is a close connection between the Hestenes method
for finding the SVD of a matrix $A$ and the Jacobi method for finding
the eigenvalues of a symmetric matrix $B = A^TA$.
Important differences are that
the formulas defining the rotation angle $\theta$ involve elements
$b_{i,j}$ of $B$ rather than inner products of columns of $A$,
and transformations must be performed on the left and right instead of just
on the right (since $(AV)^T(AV) = V^TBV$).  Instead of
permuting columns of $A$ as described in Section 6.2, we have to apply the
same permutation to both rows and columns of $B$.
An implementation on
a square systolic array of $n/2$ by $n/2$ processors is described in [\r9],
and could easily be adapted to other parallel architectures.
If less than $n^2/4$ processors are available, we can use the virtual
processor concept described in Section~2.2.

\medskip
\leftline{\bf 6.4 Other SVD and eigenvalue algorithms}
\medskip

In Section 6.2 we showed how the Hestenes method could be used to compute
the SVD of an $m$ by $n$ matrix in time $O(mn^2S/P)$ using $P = O(n)$
processors in parallel.  Here $S$ is the number of sweeps required
(conjectured to be $O(\log n)$).
In Section 6.3 we sketched how Jacobi's method could be used to compute the
eigen-decomposition of a symmetric $n$ by $n$ matrix in time $O(n^3S/P)$
using $P = O(n^2)$ processors. It is natural to ask if we can use
more than $\Omega(n)$ processors efficiently when computing the SVD.
The answer is yes -- Kogbetliantz [\r40]
and Forsythe \& Henrici [\r20] suggested an analogue of
Jacobi's method, and this can be used to compute the SVD of a {\it square}
matrix using a parallel algorithm very similar to the parallel
implementation of Jacobi's method.  The result is an algorithm which
requires time $O(n^3S/P)$ using $P = O(n^2)$ processors.
Details and a discussion of several variations on this theme
may be found in [\r10].
\medskip

In order to find the SVD of a rectangular $m$ by $n$ matrix $A$ using
$O(n^2)$ processors, we first compute the QR factorization
$QA = R$ (see Section 5.3), and then compute the SVD of the principal
$n$ by $n$ submatrix of $R$ (i.e. discard the $m-n$ zero rows of $R$).
It is possible to gain a factor of two in efficiency by preserving
the upper triangular structure of $R$ [\r48].
\medskip

The Hestenes/Jacobi/Kogbetliantz methods are not often used on a serial
computer, because they are slower than methods based on reduction to
bidiagonal or tridiagonal form followed by the QR algorithm [\r63].
Whether the fast serial algorithms can be implemented efficiently on
a parallel machine depends to some extent on the parallel architecture.
For example, on a square array of $n$ by $n$ processors it is possible to
reduce a symmetric $n$ by $n$ matrix to tridiagonal form in time
$O(n \log n)$~[\r4].  On a serial machine this reduction takes time
$O(n^3)$.  Thus, a factor $O(\log n)$ is lost in efficiency, which roughly
equates to the factor $O(S)$ by which Jacobi's method is slower than
the QR algorithm on a serial machine.  It is an open question whether the
loss in efficiency by a factor $O(\log n)$ can be avoided on a parallel
machine with $P = \Omega(n^2)$ processors.  When $P = O(n)$, ``block'' versions
of the usual serial algorithms are attractive on certain architectures [\r17],
and may be combined with the ``divide and conquer'' strategy [\r18].
Generally, these more complex algorithms are attractive on shared memory
MIMD machines with a small number of processors, while the simpler algorithms
described above are attractive on systolic arrays and
SIMD machines.

\bigskip
\leftline{\medbf References}
\medskip

\item{\r1.}	G.~Amdahl, ``Validity of the single-processor approach
	to achieving large-scale computer capabilities'',
	{\it Proc.\ AFIPS Conference} 30 (1967), 483-485.

\item{\r2.}	M.~Annaratone, E.~Arnould, T.~Gross, \htk,
	M.~Lam, O.~Menzilcioglu and J.~A.~Webb,
	``The Warp computer: architecture, implementation and
	performance'',
	\IEEETC, C-36 (1987), 1523-1538.

\item{\r3.}	C.~Bischof and C.~Van~Loan, ``The WY representation for
	products of Householder matrices'', \SISSC\ 8 (1987), s2-s13.

\item{\r4.}	\awb\ and \rpb,
	``Tridiagonalization of a symmetric matrix on a square array of
	mesh-connected processors'', {\it J.~Parallel and Distributed Computing}
	2 (1985), 261-276.

\item{\r5.}	\awb, \rpb\ and \htk,
	``Numerically stable solution of dense systems of linear equations
	using mesh-connected processors'', \SISSC\ 5 (1984), 95-104.

\item{\r6.}	S.~Borkar, R.~Cohn, G.~Cox, S.~Gleason, T.~Gross,
	\htk, M.~Lam, B.~Moore, C.~Peterson, J.~Pieper,
	L.~Rankin, P.~S.~Tseng, J.~Sutton, J.~Urbanski and J.~Webb,
	``iWarp: An integrated solution to high-speed parallel
	computing'', {\it Proc.\ Supercomputing 1988 Conference},
	Orlando, Florida, Nov.~1988, 330-339.

\item{\r7.}	\rpb, ``Parallel algorithms for digital signal processing'',
	in [\r31], 93-110.

\item{\r8.}	\rpb, \htk\ and \ftl, ``Some linear-time algorithms for
	systolic arrays'',
	in {\it Information Processing 83} (edited by R.E.A. Mason),
	North-Holland, Amsterdam, 1983, 865-876.

\item{\r9.}	\rpb\ and \ftl, ``The solution of singular-value and symmetric
	eigenvalue problems on multiprocessor arrays'', \SISSC\ 6 (1985), 69-84.

\item{\r10.}	\rpb, \ftl\ and C.~F.~Van~Loan, ``Computation of the singular
	value decomposition using mesh-connected processors'',
	{\it J.~of VLSI and Computer Systems} 1, 3 (1983-1985), 242-270.

\item{\r11.}	T.~F.~Chan, ``An improved algorithm for computing the
	singular value decomposition'', {\it ACM Trans.\ Math.\ Software}
	8 (1982), 72-83.

\item{\r12.}	E.~Chu and A.~George, 	%
	``Gaussian elimination with partial
	pivoting and load balancing on a multiprocessor'',
	{\it Parallel Computing} 5 (1987), 65-74.	%

\item{\r13.}	E.~Chu and A.~George, ``QR factorization of a dense matrix
	on a hypercube multiprocessor'', \SISSC\ 11 (1990), 990-1028.

\item{\r14.}	W.~J.~Dally and C.~L.~Seitz, ``Deadlock free message routing
	in multiprocessor interconnection networks'',
	\IEEETC\ C-36 (1987), 547-553.		%

\item{\r15.}	J.~J.~Dongarra, ``Performance of various computer using
	standard linear equations software'', Report CS-89-05,
	Computer Science Department, University of Tennessee
	(available from netlib).	%

\item{\r16.}	J.~J.~Dongarra, F.~G.~Gustavson and A.~Karp,
	``Implementing linear algebra algorithms for dense matrices on a
	vector pipeline machine'', {\it SIAM Review} 26 (1984), 91-112.

\item{\r17.}	J.~J.~Dongarra, S.~J.~Hammarling and D.~C.~Sorensen,
	``Block reduction of matrices to condensed forms for eigenvalue
	computations'', {\it LAPACK Working Note \#2},
	MCS Division, Argonne National Labs., Argonne,
	Illinois, Sept.\ 1987.

\item{\r18.}	J.~J.~Dongarra and D.~C.~Sorensen, ``A fully parallel
	algorithm for the symmetric eigenproblem'',
	\SISSC\ 8 (1987), 139-154.	%

\item{\r19.} 	M.~J.~Flynn, ``Some computer organizations and their
	effectiveness'',
	\IEEETC, C-21 (1972), 702-706.

\item{\r20.}	G.~E.~Forsythe and P.~Henrici, ``The cyclic Jacobi method for
	computing the principal values of a complex matrix'',
	{\it Trans.\ Amer.\ Math.\ Soc.\ }94 (1960), 1-23.

\item{\r21.}	G.~C.~Fox, M.~A.~Johnson, G.~A.~Lyzenga, S.~W.~Otto,
	J.~K.~Salmon and D.~W.~Walker,
	{\it Solving Problems on Concurrent Processors,
	Vol. I: General Techniques and Regular Problems}, 
	Prentice-Hall, Englewood Cliffs, New Jersey, 1988.

\item{\r22.}	G.~C.~Fox, S.~W.~Otto and A.~J.~G.~Hey,
	``Matrix algorithms on a hypercube I: matrix multiplication'',
	{\it Parallel Computing} 4 (1987), 17-31.

\item{\r23.}	K.~A.~Gallivan, R.~J.~Plemmons and A.~H.~Sameh,
	``Parallel algorithms for dense linear algebra computations'',
	{\it SIAM Review} 32 (1990), 54-135.	%

\item{\r24.}	G.~A.~Geist and M.~Heath, ``Matrix factorization on a
	hypercube'', in [\r35], 161-180.	%

\item{\r25.}	G.~A.~Geist and C.~H.~Romine, ``LU factorization
	algorithms on distributed-memory multiprocessor architectures'',
	\SISSC\ 9 (1988), 639-649.	%

\item{\r26.}	W.~M.~Gentleman, ``Least squares computations by Givens
	transformations without square roots'',
	{\it J.\ Inst.\ Math.\ Appl.\ } 12 (1973), 329-336.

\item{\r27.}	W.~M.~Gentleman and \htk, ``Matrix triangularization by
	systolic arrays'', {\it Proc.\ SPIE, Volume 298, Real-Time Signal
	Processing IV}, \SPIE, 1981.

\item{\r28.}	G.~H.~Golub and W.~Kahan, ``Calculating the singular values
	and pseudo-inverse of a matrix'', {\it J.\ SIAM Ser.\ B:
	Numer.\ Anal. }2 (1965), 205-224.

\item{\r29.}	G.~H.~Golub and \ftl, ``Singular value decomposition:
	applications and computations'', ARO Report 77-1, 
	{\it Trans.\ 22nd Conference of Army Mathematicians}, 1977, 577-605.

\item{\r30.}	G.~H.~Golub and C.~Reinsch, ``Singular value decomposition
	and least squares solutions'', {\it Numer.\ Math.\ }14 (1970), 403-420.

\item{\r31.}	G.~H.~Golub and P.~Van~Dooren, {\it Numerical Linear Algebra,
	Digital Signal Processing and Parallel Algorithms},
	Springer-Verlag, 1990, 93-110.

\item{\r32.}	G.~H.~Golub and C.~Van~Loan, {\it Matrix Computations},
	Johns Hopkins Press, Baltimore, Maryland, 1983.

\item{\r33.}	J.~L.~Gustafson, G.~R.~Montry and R.~E.~Benner,
	``Development of parallel methods for a 1024-processor hypercube'',
	\SISSC\ 9 (1988), 609-638.	%

\item{\r34.}    E.~R.~Hansen, ``On cyclic Jacobi methods'',
        {\it J.\ SIAM} 11 (1963), 448-459.

\item{\r35.}	M.~T.~Heath (editor), {\it Hypercube Multiprocessors 1986},
	SIAM, Philadelphia, 1986.

\item{\r36.}	D.~Heller, ``A survey of parallel algorithms in numerical
	linear algebra'', {\it SIAM Review} 20 (1978), 740-777.

\item{\r37.}	M.~R.~Hestenes, ``Inversion of matrices by biorthogonalization
	and related results'', {\it J.\ SIAM} 6 (1958), 51-90.

\item{\r38.}	N.~J.~Higham, {\it Exploiting fast matrix multiplication
	within the level 3 BLAS}, Report TR 89-984, Dept. of Computer Science,
	Cornell University, 1989.	%

\item{\r39.}	M.~Ishii, G.~Goto and Y.~Hatano, ``Cellular array processor
	CAP and its application to computer graphics'',
	{\it Fujitsu Sci.\ Tech.~J.\ } 23 (1987), 379-390.

\item{\r40.}	E.~Kogbetliantz, ``Diagonalization of general complex matrices
	as a new method for solution of linear equations'',
	{\it Proc.\ Internat.\ Congress of Mathematicians}, Amsterdam,
	Vol. 2, 1954, 356-357.

\item{\r41.}	\htk, ``Why systolic architectures ?'',
        {\it IEEE Computer} 15, 1 (1982), 37-46.

\item{\r42.}	\htk\ and C.~E.~Leiserson, ``Systolic arrays (for VLSI)'',
	{\it Sparse Matrix Proc.\ 1978}, SIAM, Philadelphia, 1979, 256-282.

\item{\r43.}	\htk, R.~F.~Sproull and G.~L.~Steele, Jr.~(eds.),
	{\it VLSI Systems and Computations}, Computer Science Press, 1981.

\item{\r44.}	S.~Y.~Kung (editor), {\it VLSI and Modern Signal Processing},
	Proceedings of a Workshop held at Univ.\ of Southern California,
	Nov.\ 1982.

\item{\r45.}	S.~Y.~Kung, {\it VLSI Array Processors},
	Prentice-Hall International, 1988.

\item{\r46.}	G.~Li and T.~F.~Coleman, ``A new method for solving
	triangular systems on distributed-memory message-passing
	multiprocessors'', \SISSC\ 10 (1989), 382-396.	%

\item{\r47.}	\ftl, ``A rotation method for computing the QR-decomposition'',
	\SISSC\ 7 (1986), 452-459.

\item{\r48.}	\ftl, ``A triangular processor array for computing singular
	values'', {\it J.\ of Linear Algebra and its Applications} 77 (1986),
	259-273.

\item{\r49.}	\ftl, ``Architectures for computing eigenvalues and SVDs'',
	{\it Proc.\ SPIE, Vol.\ 614, Highly Parallel Signal Processing
	Architectures}, 1986, 24-33.

\item{\r50.}	\ftl\ and H.~Park, ``On parallel Jacobi orderings'',
	\SISSC\ 10 (1989), 18-26.

\item{\r51.}	\ftl\ and H.~Park, ``A proof of convergence for two parallel
	Jacobi SVD algorithms'', \IEEETC\ 30 (1989), 806-811.

\item{\r52.}	C.~Moler, ``Matrix computations on distributed memory
	multiprocessors'', in [\r35], 181-195.	%

\item{\r53.}	W.~Moore, A.~McCabe and R.~Urquhart (editors),
	{\it Systolic Arrays}, Adam Hilger, Bristol, 1987.

\item{\r54.}	A.~Pothen and P.~Raghavan, ``Distributed orthogonal
	factorization: Givens and House\-holder algorithms'',
	\SISSC\ 10, 6 (1989), 1113-1134.

\item{\r55.}	A.~H.~Sameh and D.~J.~Kuck, ``On stable parallel linear
	system solvers'', \JACM\ 25 (1978), 81-91.

\item{\r56.}	D.~E.~Schimmel and \ftl, ``A new systolic array for the
	singular value decomposition'', {\it Proc.\ Fourth MIT Conference
	on Advanced Research in VLSI}, 1986, 205-217.

\item{\r57.}	R.~Schreiber and C.~Van~Loan, ``A storage-efficient WY
	representation for products of Householder transformations'',
	\SISSC\ 10 (1989), 53-57.

\item{\r58.}	H.~Schr\"oder,
	``The instruction systolic array~-- A tradeoff between
	flexibility and speed'',
	{\it Computer Systems Science and Engineering} 3 (1988), 83-90.

\item{\r59.}	G.~Shroff and R.~Schreiber, {\it On the convergence of the
	cyclic Jacobi method for parallel block orderings},
	Report 88-11, Department of Computer Science, Rensselaer Polytechnic
	Institute, Troy, New York, May 1988.

\item{\r60.}	G.~W.~Stewart, {\it Introduction to Matrix Computations},
	Academic Press, New York, 1973.

\item{\r61.}	G.~W.~Stewart, ``A Jacobi-like algorithm for computing the
	Schur decomposition of a non-Hermitian matrix'',
	\SISSC\ 6 (1985), 853-864.

\item{\r62.}	G.~W.~Stewart, ``A parallel implementation of the QR
	algorithm'', {\it Parallel Computing} 5 (1987), 187-196.

\item{\r63.}	J.~H.~Wilkinson, {\it The Algebraic Eigenvalue Problem},
	Clarendon, Oxford, 1965.

\vfill\eject
\bye